\documentclass[aps,prc,twocolumn,showpacs,superscriptaddress, twoside,showpacs]{revtex4-1}
\usepackage{color,hyperref}
\hypersetup{colorlinks,breaklinks,linkcolor=blue,urlcolor=blue,anchorcolor=blue,citecolor=blue}
\usepackage{CJK}
\usepackage{graphicx}
\usepackage{dcolumn}
\usepackage{bm}
\usepackage{booktabs}
\usepackage{amsmath}
\usepackage{color}

\begin{document}

\title{Smooth band termination in $^{115}$I}

\author{C.~ M.~Petrache}
\affiliation{Universit\'e Paris-Saclay, CNRS/IN2P3, IJCLab, 91405 Orsay, France}
\affiliation{Institute of Modern Physics, Chinese Academy of Sciences, Lanzhou 730000, China}
\author{I. Ragnarsson}
\affiliation{Division of Mathematical Physics, LTH, Lund University, P.O. Box 118, SE-221 00 Lund, Sweden}

%----------------------------------------------------------------------------------------------------
\begin{abstract}

The cranked Nilsson-Strutinsky formalism is employed to investigate the band structure of $^{115}$I from low to high spin. A set of eight rotational bands with well established spins and parities are analyzed. Five bands start at low spin and are assigned to the single-particle proton orbitals ($d_{5/2}$, $g_{7/2}$), $g_{9/2}$, and $h_{11/2}$, which are close to the Z=53 Fermi surface for a deformation of $\varepsilon_2=0.2$. The analysis of the observed level energies relative to a rotating liquid drop and spins versus $\gamma$-ray energies allow to track the evolution of the configurations with increasing spin. It is of special interest that one band in the valence space is observed to a smooth termination at $I_{max} = 67/2^+$.

%The sharp backbendings observed in the bands built on the $\nu d_{5/2}$ and $\nu g_{7/2}$ orbitals at $I\approx10$ are interpreted as  $\nu (h_{11/2})^2$ alignments, while the backbendings at $I\approx20$ are interpreted as crossings between configurations involving two and three $h_{11/2}$ neutrons.

\end{abstract}

%\pacs{21.10.Re, 21.60.Ev, 23.20.Lv, 27.60.+j}

%\keywords{ Nuclear reaction:$^{58}$Ni($^{64}$Zn,$\alpha3p$)^{115}I$; E= 275 MeV; Measured  $\gamma$-$\gamma$-$\gamma$-coincidences;  E$_\gamma$; I$_\gamma$; angular distributions; polarization; $^{115}$I deduced levels; spin and parity; model calculation}

%\maketitle
\maketitle

\section{Introduction}

The band structures of deformed nuclei with a reduced number of protons and neutrons above the $Z,N=50$ magic gaps offer optimum test benches for theoretical models conceived to describe the interplay between collective and single-particle motion. One of these models is cranked Nilsson-Strutinsky (CNS) \cite{Bengtsson1985},
developed in particular for the interpretation of high-spin states. As pairing is neglected in this model, it has generally not been applied to low-spin states. However, the spin dependence of the difference between observed and calculated energies should mirror the pairing energy. Indeed, if it is accepted that the difference between experiment and calculations can be understood as an average pairing energy, it appears that the model can be used for all spin values, see e.g. \cite{Rag24}.  An exception may be the region of  the first band-crossing in prolate nuclei where the Fermi level is located at the bottom of a high-$j$ shell. However, the general features of such band-crossings are well understood from
simple models, see e.g. \cite{Steph73} or \cite{NR}.

%{\it Out: and therefore seldom used to interpret bands that develop from low to very high spin.  The pairing being neglected, the experimental low-spin part of the rotational bands is overestimated. But when one or more pairs of nucleons are broken, or when particle-hole excitations are involved, the pairing is significantly reduced and the single-particle approximation can be applied. In such circumstances the CNS formalism becomes a powerful tool for the investigation of rotational bands since it allows to track a given configuration through crossings and also unveil the shape evolution with increasing spin. }

The present work is devoted to $^{115}$I, a nucleus studied in the past using $\gamma$-ray spectrometers with low and moderate efficiency \cite{115i-Piel,115i-Paul-Stony-Brook,115i-Paul-Chalk-River,115i-Moon}, but also very recently in a high-statistics experiment with the JUROGAM array \cite{jurogam3}, which allowed to extend the previously known bands and to identify several bands to spin values above $I$=30 \cite{115I-exp}. The use of clover detectors in the JUROGAM experiment was beneficial, allowing the firm
assignment of spins and parities to all bands from angular correlations and polarization measurements. The spins and parities of several levels reported previously have been changed, and the level scheme has been largely revised. The experimental information on the $\gamma$-ray transitions observed in $^{115}$I will be reported in a forthcoming paper \cite{115I-exp}. In the present work we only show the bands relative to a rotating liquid drop energy in Fig. \ref{fig1}, and give the band-head spins and parities, the band-head energies, and the transition energies of the bands in Table \ref{table1}. 

\begin{table*}[ht!]
\centering
\caption{Experimental information for the bands of $^{115}$I discussed in the present work: band labels, band-head spins and parities, band-head energies (in keV), parities of the bands, and transition energies (in keV) of the in-band transitions. The new transitions, to be published in a forthcoming paper \cite{115I-exp}, are indicated wtih bold characters. For band 8, the two cascades of $E2$ transitions interconnected by dipole transitions are labeled 8e and 8o, corresponding to even and odd signature, respectively. The highest transitions of band 7 are from Ref. \cite{115i-Paul-Chalk-River}. The energies of these transitions have not been observed in Ref. \cite{115I-exp}, therefore the energies are in parentheses. } 
 \vspace{0.5mm}
\label{table1}
%\begin{ruledtabular}
%\renewcommand\arraystretch{0.3}
\begin{tabular}{llllllllll}
~~1                  & ~~2          &  ~~3          &  ~~4       & ~~5          & ~~6          &  ~~7         &   ~8e         &  ~8o         \\
$11/2^-$           & $43/2^-$  & $17/2^+$  & $9/2^+$  & $7/2^+$   & $7/2^+$   & $43/2^+$  & $9/2^+$     &  $11/2^+$\\
~835                & 6479        & 2409         & 632         & 599        & ~56          & 7096         & 563           &  876         \\
~~$-$               & ~~$-$      & ~~+           & ~+           & ~~+          & ~~+          & ~~+          & ~~+           & ~~+         \\
\hline
     ~410           & 1032        & ~505         &  617        &$\bf~654$ &  ~736       & $\bf1028$ & ~633         & ~662        \\
     ~516           & 1053        & ~488         &  705        &$\bf~762$ &  ~845       & $\bf1090$ & ~700         & ~732        \\   
     ~622           & $\bf1088$& ~646         &  781        &$\bf~710$ &  ~930       & $\bf1146$ & ~758         & ~778        \\  
     ~728           &                 & ~746         &  781        &~609          &  ~600      & $\bf1206$ & ~787         & ~788        \\
     ~796           &                 & ~820         &  665        &~288          &  ~593      & $\bf1298$ & ~797         & $\bf~823$ \\
     ~849           &                 & ~924         &  818        &~678          &  ~686      & $\bf1382$ & $\bf~862$ & $\bf~905$ \\
     ~893           &                 & $\bf~924$ &                &~866          &  ~786      & $\bf1467$ & $\bf~938$ & $\bf~977$ \\
     1015           &                 & $\bf~736$ &                &$\bf1091$  &  ~880      & $\bf1575$ & $\bf1010$  & $\bf1035$ \\  
     1055           &                 & $\bf~993$ &               &$\bf1285$   &  ~936      & $\bf1705$ & $\bf1071$  & $\bf1092$ \\
     1127           &                 & $\bf1034$ &               &                   &  ~934       &  (1860)     &                   &                  \\   
     1211           &                 & $\bf1039$ &               &                   & $\bf~704$ &  (2025)    &                   &                  \\  
     1319           &                 & $\bf1715$ &               &                   & $\bf1155$ &                 &                   &                  \\
     $\bf1382$   &                 &                   &               &                  & $\bf1218$ &                 &                   &                  \\
     $\bf1429$   &                 &                   &               &                  & $\bf1552$ &                 &                   &                  \\
     $\bf1652$   &                 &                   &               &                  & $\bf2004$ &                 &                   &                  \\
                        &                 &                   &               &                  & $\bf2470$ &                 &                   &                    
    \end{tabular}
%\end{ruledtabular}
\end{table*}

\section{General considerations}

The band structure of $^{115}$I analyzed in the present work is composed of two negative-parity bands (bands 1 and 2) and six positive-parity bands (bands 3$-$8). Band 1 with negative parity and four positive-parity bands start at low spin and are interpreted as built on one-quasiparticle configurations.  Band 1 built on the $11/2^-$ state is clearly based on the $\nu h_{11/2}$ configuration. Band 3, previously reported with negative-parity \cite{115i-Moon}, has in fact positive parity, as established from the polarization asymmetry of the 936-keV transition to the $27/2^-$ state of band 1, which  indicates an $E1$ character  \cite{115I-exp}. It decays via five transitions to states of the negative-parity band 1, like in the isotone $^{117}$Cs \cite{117cs-jodidar}, and to the positive-parity bands 4 and 6. Numerous interconnecting transitions exist between the four bands 3$-$6, strongly supporting similar configurations.
%, which should be based on the $\nu d_{5/2}$ and $\nu g_{7/2}$ orbitals assigned to the  $5/2^+$ and $7/2^+$ lowest excited states of $^{115}$I. Bands 3 and 4 have positive signature, and bands 5 and 6 have negative signature, suggesting configurations based on the four orbitals %with (parity, signature)=($\pi$=+, $\alpha$=$\pm$1/2) arising from the splitting of the $1/2^+[420]$ and $3/2^+[422]$ Nilsson orbitals at non-zero rotational frequency. 
%at rotational frequencies above $\hbar\omega$$\approx$0.3 MeV, corresponding to the energies of the lowest transitions in each band. 
%To our knowledge, this is the only case in this mass region where bands  built on all four $1/2^+[420]$($\alpha$=$\pm$1/2) and  $3/2^+[422]$($\alpha$=$\pm$1/2) orbitals are observed. 
Band 8 consisting of two nearly degenerate cascades of $E2$ transitions interconnected by  $M1/E2$ dipole transitions is clearly based on a configuration with one hole in the $\pi g_{9/2}$ orbital, which give rise to strongly coupled bands, like in several nuclei of the $A$=115  mass region, see e. g.  \cite{117cs-jodidar,evans,117I-warring,Sta01,119cs-zheng,114i-jodidar}.

Once the configurations are assigned to the low-spin parts of the bands, one would like to understand their evolution at high spin. One can get a first idea by realizing the resemblance of the observed bands with those recently reported in the $^{117}$Cs isotone  \cite{117cs-jodidar}, in which the low-spin bands 1, 2, 3, 4 correspond one-to-one to the bands 1, 3, 4, 8 of $^{115}$I. This is not surprising, because the two nuclei are isotones and the number of protons is different by two.
%even though the deformations of the two nuclei can be a bit different. 
%\subsection{Energies relative to a rotating liquid drop and spin versus $\gamma$-ray energy}

\begin{figure}[!htb]
\hskip -. cm
\vskip -. cm
\rotatebox{-0}{\scalebox{0.47}{\includegraphics[trim=0 0 0 0, width=\textwidth]{fig1a}}}
\rotatebox{-0}{\scalebox{0.57}{\includegraphics[trim=0 0 0 0, width=\textwidth]{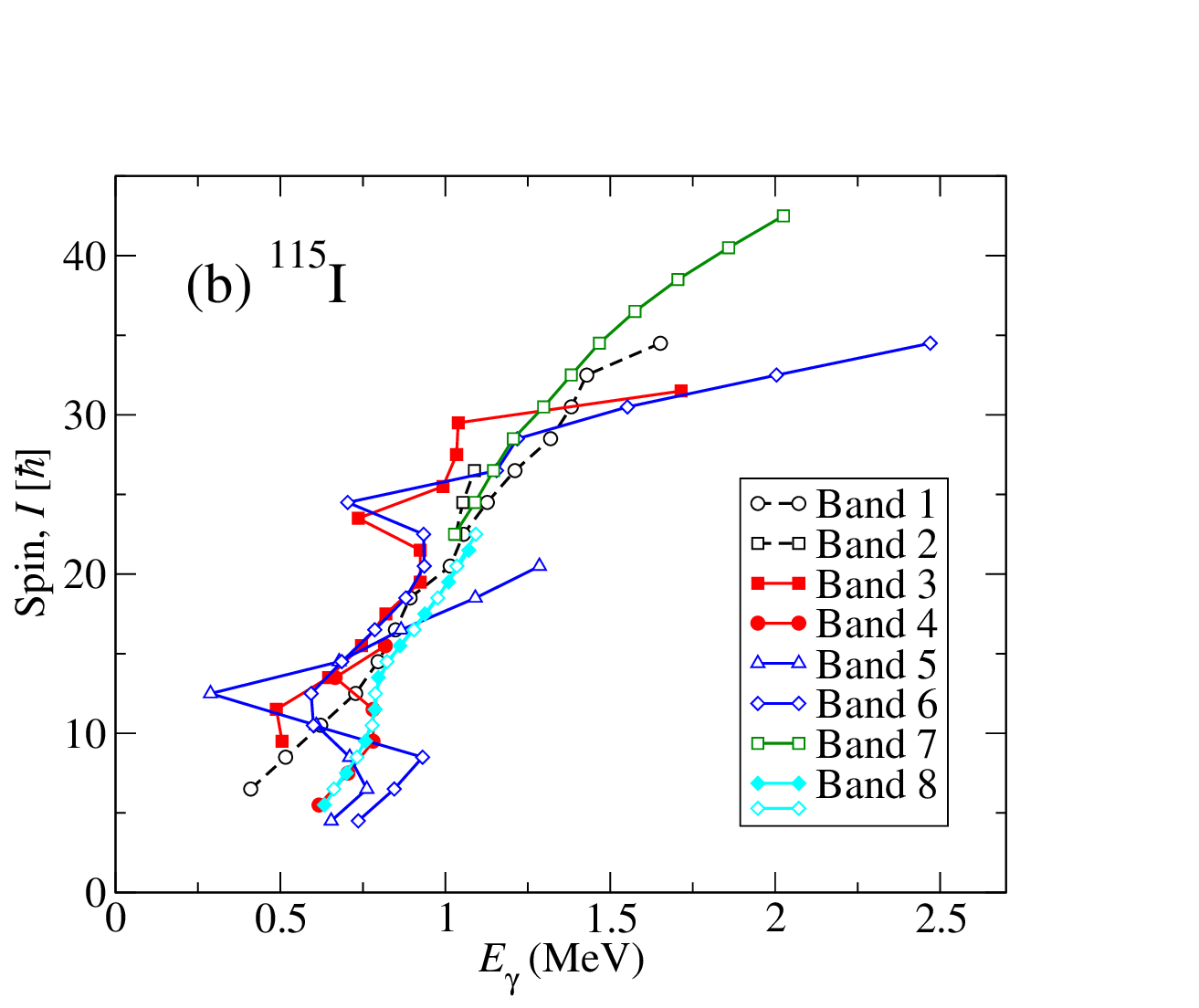}}}
\vskip -0.3 cm
\caption {\label{fig1} (a)  $E$$-$$E_{rld}$ values for the bands of $^{115}$I. (b)  Spin $I$ versus $\gamma$-ray energy $E_{\gamma}$. Note that bands 4, 5 and 6 show a clear backbend at $E_{\gamma}\approx 0.7$ MeV, while only a minor upbend is seen for band 1.}
\vskip -0.2cm
\end{figure}

To get a first idea of the evolution of the bands with increasing spin, we analyzed the energies relative to a rotating liquid drop $E$$-$$E_{rld}$ as a function of spin $I$, and the spin $I$ versus the $\gamma$-ray energy $E_{\gamma}$, which are shown in Fig. \ref{fig1}.  In the $E$$-$$E_{rld}$ plots one can see that all bands exhibit up-sloping patterns at low spin, which are caused by the pairing correlations. Note however, that the steep upsloping behavior is broken already at $I \approx 10$, suggesting a drastic reduction of the pairing at higher spin.

%One can also see that bands 1 and 8 have different behaviors than bands 3$-$6: bands 1 and 8 exhibit smooth curves, with gradually changing slopes, while bands 3$-$6 exhibit drastic slope changes at $I$$\approx$10 and $I$$\approx$20.  
One can distinguish three spin intervals in which bands 3$-$6 have regular behavior: $I$$\approx$0$-$10, $I$$\approx$10$-$20 and $I$$>$20. For $I$$\approx$0$-$10 all curves are upsloping, with similar slope for bands 3$-$6 and 8, and a definitely smaller slope for band 1, which makes it yrast up to $I$=39/2, where band 2 becomes yrast. For $I$$\approx$10$-$20 bands 3 and 6 behave like signature partners, while bands 4 and 5 have different curvatures, but lie at excitations not much different from those of bands 3 and 6.

%%%%%%%
One can also consider that bands 4 and 5 are signature partners, as one expects the existence of four bands built on both signatures of the $(gd)$ orbitals. In fact, the energy spacing of the four [420]1/2($\alpha$=$\pm$1/2) and [422]3/2($\alpha$=$\pm$1/2) proton orbitals depends on the deformation and on the rotational frequency. For deformations of $\varepsilon_2$$\approx$0.2, $\gamma$=$0^{^\circ}$ and frequencies of $\hbar\omega$$\approx$0.3 MeV, the $(gd)$ orbitals with positive signature are lowest and highest in energy, see Fig. \ref{fig2}, which is in qualitative agreement with the $E$$-$$E_{rld}$ plots, and offer an explanation of the non-observation of band 3 at low spin where it is too excited. 
%e. g. Fig. 8 in Ref. \cite{116I-moon} drawn for $^{116}$I but also approximately valid for $^{115}$I

%%%%%%%
For $I$$>$20, the slopes of bands 3 and 6 change drastically, which makes them rapidly yrast. A special attention should be given to band 6, which becomes rapidly non yrast at the two highest states, and behaves like a smooth terminating band \cite{IR-1995,ing-phys-rep}.

%After the first crossing at $I$$\approx$10 in bands 3$-$6 and at $I$$\approx$18 in band 1, the $E$$-$$E_{rld}$ values are down-sloping, as in the CNS calculations, revealing a significant reduction of the pairing correlations. Bands 4$-$6 are observed starting from spins well below the crossing at $I$$\approx$10 and have similar slopes, while band 3 is observed starting from two states below the crossing at $I$$\approx$10, suggesting a higher excitation at low spin than bands 4$-$6. 

%Another feature present in the  $E$$-$$E_{rld}$ plots is the yrastness of bands 2, 3 and 6 at spins higher than $I$$\approx$20. This can be attributed to the crossing with configurations which become yrast at high spin if more $h_{11/2}$ neutrons are involved in the configurations, two more in band 2 relative to band 1, and one more in bands 3 and 6 relative to the configurations below the crossing at $I$$\approx$20. 

The $I$ versus $E_{\gamma}$ plots in Fig. \ref{fig1}(b) show gradual alignments in bands 1 and 8, built on the $\pi h_{11/2}$ and $\pi g_{9/2}$ orbitals, respectively. This suggests a $\nu (h_{11/2})^2$ alignment in band 1, where the $\pi (h_{11/2})^2$ alignment is not possible because only one $h_{11/2}$ proton in present in the low-spin configurations. 
%This is different from the alignment observed in the corresponding bands in the $^{117}$Cs isotone \cite{117cs-jodidar}, where three $h_{11/2}$ protons are present in the low-spin configurations, and the $\pi (h_{11/2})^2$ alignment is possible. 
On the contrary, bands 3$-$6 built on the $(gd)$ orbitals exhibit sharp alignments at $I$$\approx$10 and $I$$\approx$22.
%, which suggests a lower deformation and larger softness of the $^{115}$I core relative to that of $^{117}$Cs. Similar sharp crossings are also observed at $I$$\approx$22 in bands 3 and 6 of $^{115}$I, which are different from the smooth alignments observed in the corresponding bands of $^{117}$Cs. 
%Second crossings are observed at $I$$\approx$22 in bands 3 and 6, which cannot be due to the $\pi (h_{11/2})^2$ alignment, and therefore have to be induced by crossings with other configurations which become yrast at high spin. 
In order to understand the nature of the observed crossings, we performed CNS calculations from which we could estimate the deformation and the excitation of the configurations close to yrast in the different spin regions, as will be presented in the  following.

\hfill \break

\subsection{CNS calculations and configuration assignments}

In the CNS formalism \cite{Bengtsson1985,Car06}, the nucleus rotates about one of its principal axes and pairing is neglected. The deformation is optimized in $(\varepsilon_2, \gamma, \varepsilon_4)$ for each configuration at each spin value. The configurations are labelled by the number of particles in low-$j$ and high-$j$ orbitals, in the different ${\cal N}$-shells. For $^{115}$I, they can be defined relative to a $^{100}$Sn core as,
\[
\pi(g_{9/2})^{-p_1}(dg)^{p_2}(h_{11/2})^{p_3} \nu[(dg)(sd)]^{n_1}(h_{11/2})^{n_2},
\]
for which the short-hand notation $[(p_1) p_2 p_3;n_1 n_2 ]$ is used. When $p_1 = 0$, it is not written out. Mainly depending on deformation, the pseudo-spin partners $d_{5/2}g_{7/2}$ $(dg)$ and $s_{1/2}d_{3/2}$ $(sd)$ can be  distinguished in the CNS formalism in some cases but not in other cases. In the present case with $p_1 = 0$, the proton $(sd)$ orbitals will not be competitive in energy so we will mainly consider the $(dg)(h_{11/2})$ valence space with 26 orbitals  for both protons and neutrons, out of the 32 present in the 50$-$82 major shell.  Note that all particles are listed and not just those which might be considered as active. Furthermore, the labels do not refer to the pure $j$-shells, but rather to the dominating amplitudes in the orbitals of the cranking Hamiltonian. For an odd number of particles in a group, the signature is specified as a subscript $+$ for $\alpha$=$+$1/2 or $-$ for $\alpha$=$-$1/2.  We use the so-called Lund convention for the triaxiality parameter $\gamma$ in relation to the main rotation axis, where for positive $\gamma$ shape, 0$^{\circ}$$<$$\gamma$$<$60$^{\circ}$, the rotation $(x)$ axis is the shortest principal axis, while for negative $\gamma$ shape, $-60^{\circ}$$<$$\gamma$$<$0$^{\circ}$, it is the intermediate principal axis.  In the present calculations the $A=110$ parameters are used \cite{ing-phys-rep,A=120-140}.

\begin{figure}[!ht]
\hskip -. cm
\vskip -. cm
\rotatebox{-0}{\scalebox{0.48}{\includegraphics[trim=10 0 0 0, width=\textwidth]{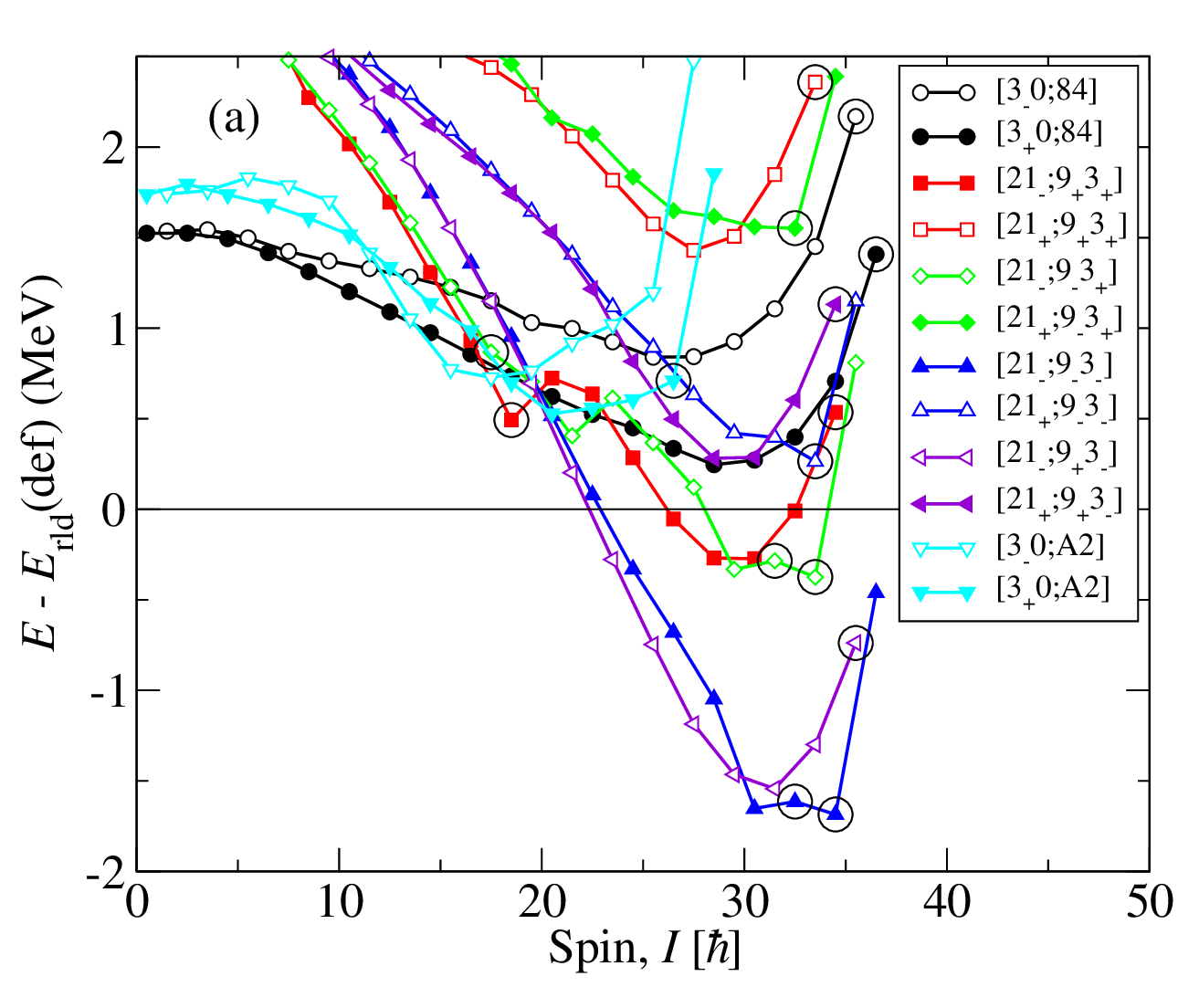}}}
\vskip 1.3 cm
\rotatebox{-0}{\scalebox{0.48}{\includegraphics[trim=0 0 0 0, width=\textwidth]{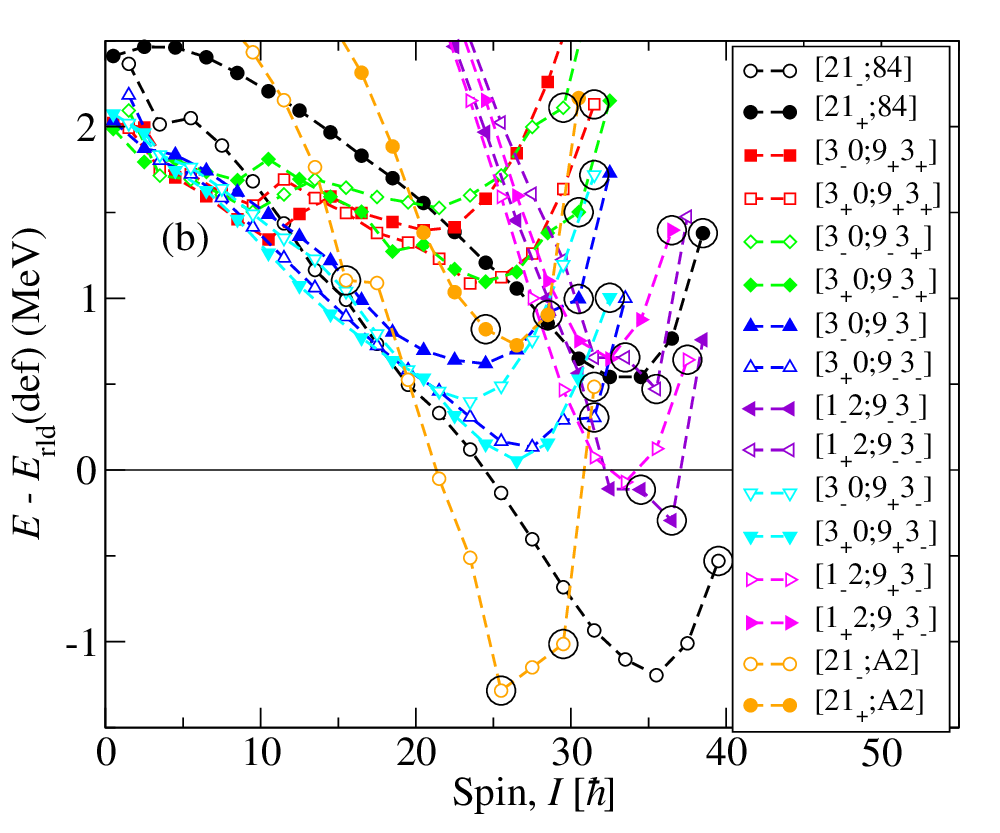}}}
\vskip -0.3 cm
\caption {\label{fig1s} The calculated low-energy CNS configurations in the valence space of $^{110}$I: (a) positive and (b) negative parity. States with positive (negative) signature are drawn
  with filled (open) symbols. }
\vskip -0.2cm
\end{figure}

Considering only valence space configurations, after calculating all configurations with different parity and signature which become yrast in the $I$=0$-$40 spin range, we concluded that the favored configurations are $[30;n_1n_2]$ and $[21;n_1n_2]$ for protons, and $[p_1p_2;84]$, $[p_1p_2;93]$, and $[p_1p_2;A2]$ for neutrons, where A is 10 in the hexadecimal system. These four proton and six neutron configurations are then combined. The $[12;n_1n_2]$ proton configurations were also analyzed, 
%to check if they might come close to yrast in energy.  In general, 
and found to be  calculated high in energy, but come closest to yrast at high spin. Thus, four configurations of this type which come down in energy for $I$$>$30 are also included. These calculated configurations are shown in Fig. \ref{fig1s}, where they are split into positive- and negative-parity configurations to avoid a too crowded figure.

\begin{figure}[!ht]
\hskip -. cm
\vskip -. cm
\rotatebox{-0}{\scalebox{0.48}{\includegraphics[trim=10 0 0 0, width=\textwidth]{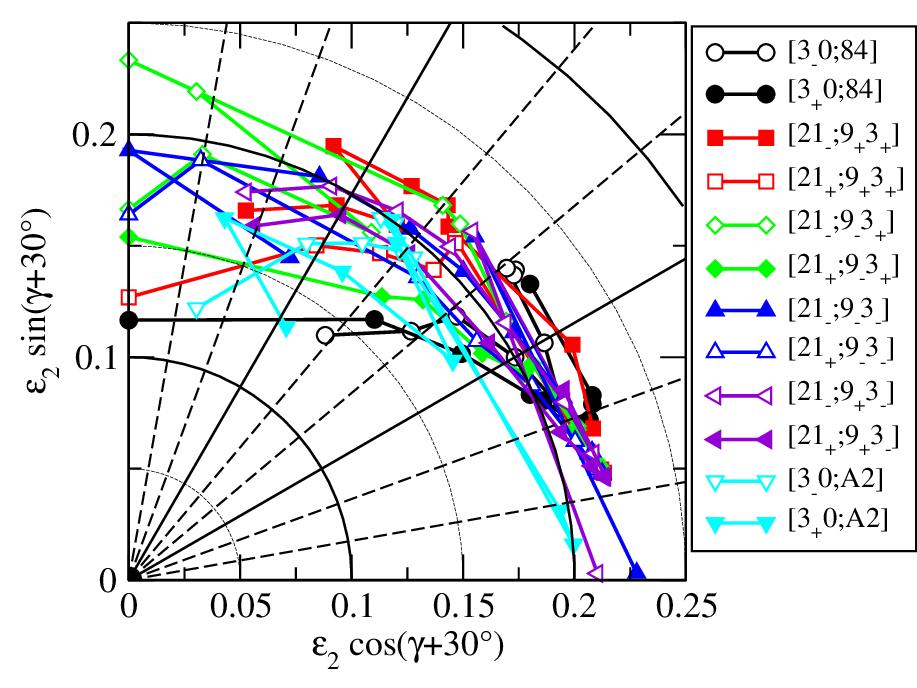}}}
\vskip -0.3 cm
\caption {\label{fig2s} Deformation trajectories for the positive-parity valence space configurations
  in Fig. 1(a).}
\vskip -0.2cm
\end{figure}

In addition to the excitation energies, it is also interesting to follow the shape changes with increasing spin of the different configurations. To get a general overview, the deformation
trajectories for all positive-parity configurations of Fig. 1(a) are shown in Fig. \ref{fig2s}. The outcome is that almost all bands start at prolate shape ($\gamma$ within $\pm15^{\circ}$), with $\varepsilon_2 \ge 0.20$.
 %slightly larger than or approximately equal 0.20. 
 With increasing spin there is a trend towards termination at $\gamma$=60$^{\circ}$, 
 %where these configurations terminate at 
 and rather different deformations in the range $\varepsilon_2$ $\approx$0.12$-$0.20. 
 
\begin{figure}[!ht]
\hskip -. cm
\vskip -. cm
\rotatebox{-0}{\scalebox{0.48}{\includegraphics[trim=10 0 0 0, width=\textwidth]{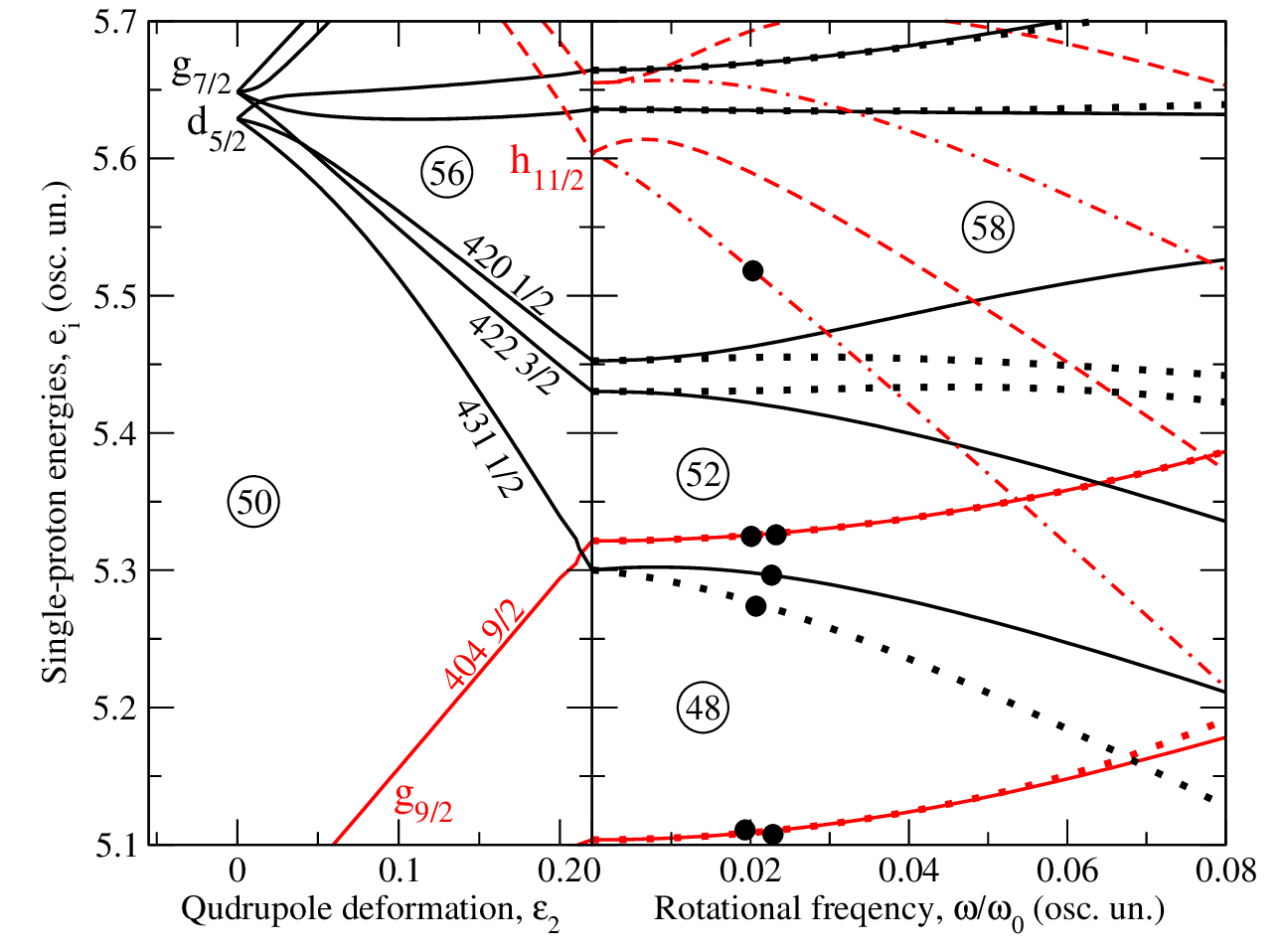}}}
\vskip 1.5 cm
\rotatebox{-0}{\scalebox{0.48}{\includegraphics[trim=10 0 0 0, width=\textwidth]{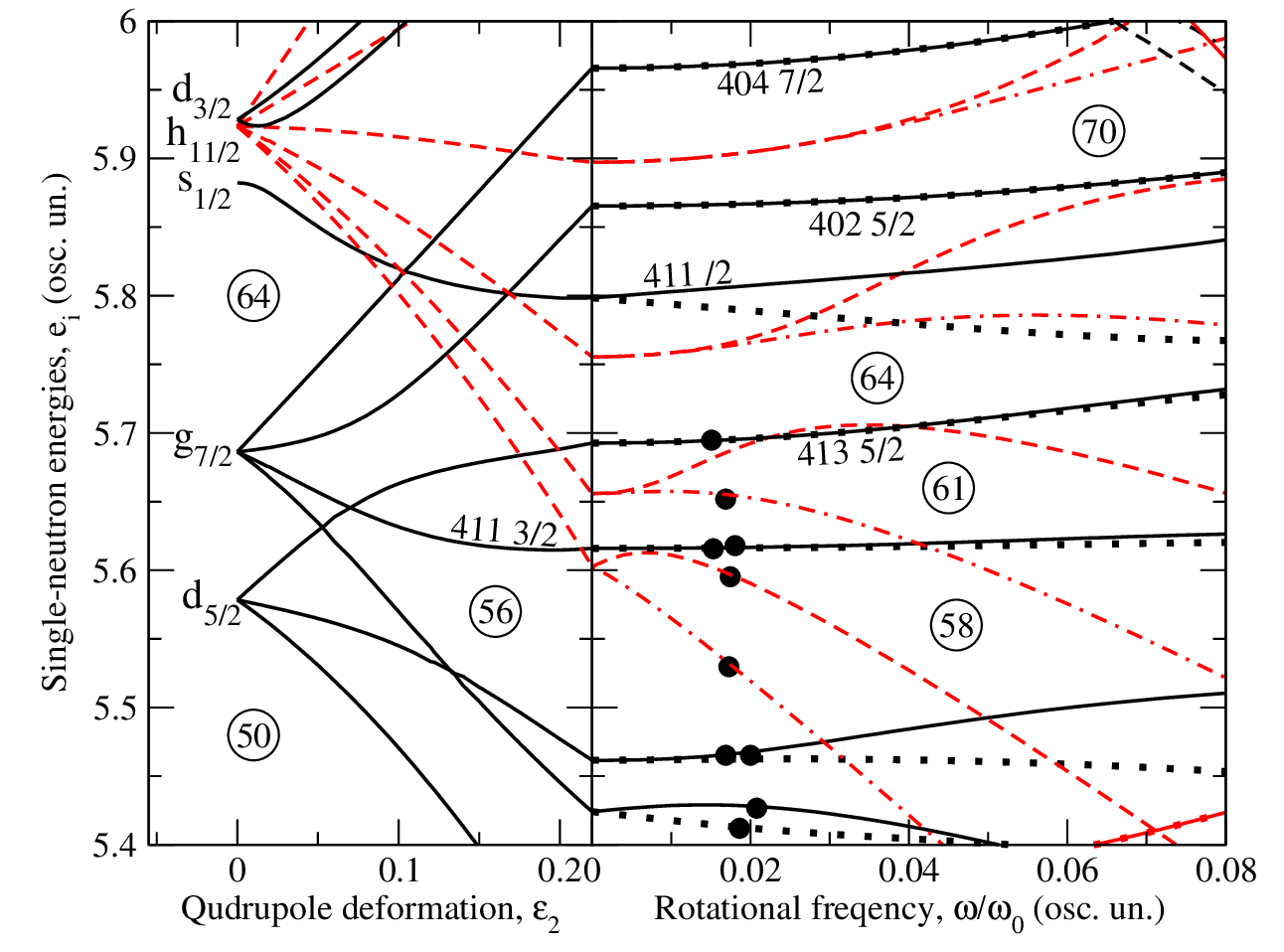}}}
\vskip -0.3 cm
\caption {\label{fig2} (a) Single-proton and (b) single-neutron Routhians at a prolate shape typical for $^{115}$I. The orbitals are traced back to spherical shape on the left.  Line convention for $(\pi,\alpha)$ orbitals: continuous for $(+,+1/2)$, dotted for $(+,-1/2)$, dashed for $(-,+1/2)$, doted-dashed for $(-,-1/2)$. The occupation of the orbitals in the terminating $[21_-;9_+3_-] $ ($\pi(dg)^2(h_{11/2})^1  \nu(dg)^9(h_{11/2})^3$) configuration is illustrated by filled black circles. } 
\vskip -0.2cm
\end{figure}

A typical low-spin deformation in $^{115}$I is ($\varepsilon_2$=0.22, $\gamma$=$0^{\circ}$), but with large fluctuations.  Thus single-neutron Routhians at this deformation are
drawn in Fig. \ref{fig2}. We can see that relative to the $N$=64 shell gap, favored configurations in $^{115}$I, which has 62 neutrons, are formed with two holes in either the $h_{11/2}$ or the $(gd)$ orbitals below the gap, or with one hole in each of these groups, leading to configurations with 2, 3 or 4 neutrons in $h_{11/2}$ orbitals. For protons, the $Z$=50 core is closed for the valence space, but at a somewhat higher deformation, configurations with holes in the $g_{9/2}$ orbitals might become competitive.
% lowest lying bands, but at somewhat higher deformation also the $g_{9/2}$ orbital can be occupied.
%Both signatures of the two lowest $(gd)$ orbitals leading
%to four orbitals at non-zero rotational frequency are clearly
%favored.
It follows that in the yrast region, the third proton above $Z$=50 can fill the two signatures of the $(gd)$, $h_{11/2}$ or $g_{9/2}$ orbitals.

\subsubsection{Bands 1 and 2}

\begin{figure}[!ht]
\hskip -. cm
\vskip -. cm
\rotatebox{-0}{\scalebox{0.46}{\includegraphics[trim=10 0 0 0, width=\textwidth]{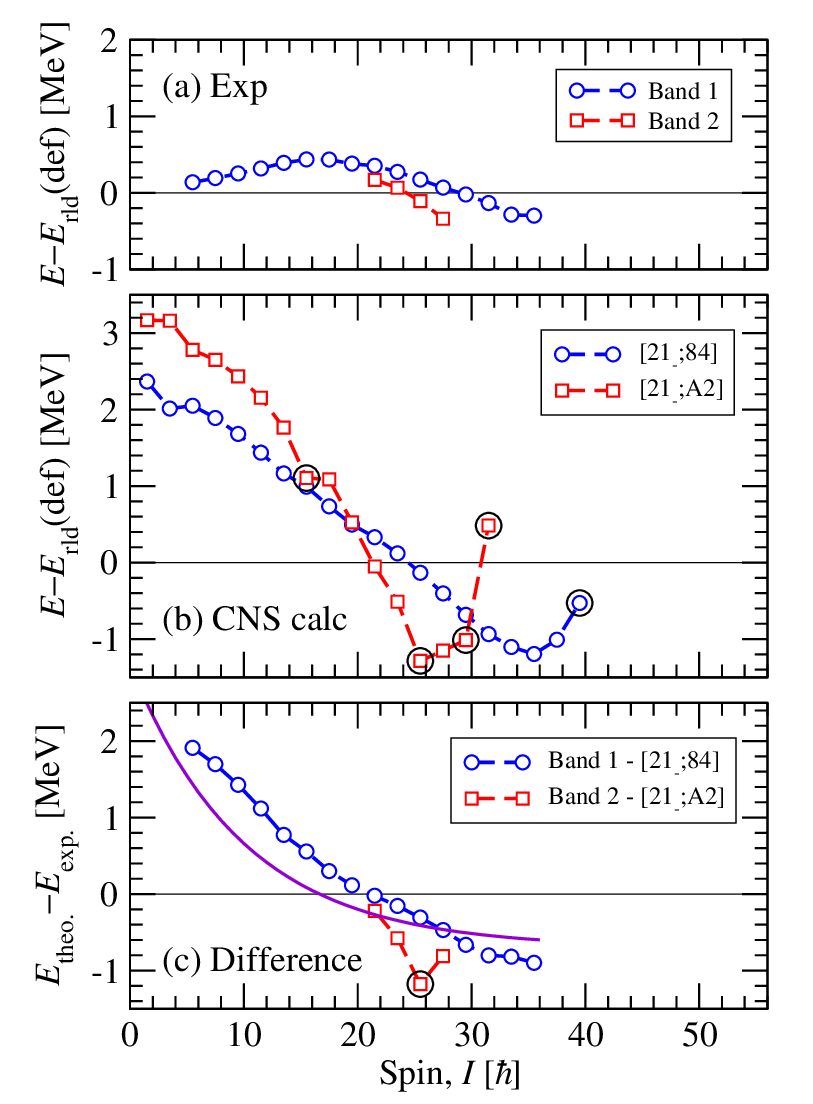}}}
\vskip -0.3 cm
\caption {\label{fig3} The observed bands 1 and 2 are drawn relative to the rotating liquid drop energy in the upper panel with the calculated configurations assigned to these bands drawn relative to the same reference in the middle panel. Note that A refers to 10 in the hexadecimal system. The difference between the observed and calculated bands are drawn in the lower panel where also a curve, -0.7 + 3.7$\times\exp{I/10}$, is drawn to indicate how the pairing energy might vary with spin. This same curve is drawn in all figures comparing observed and calculated bands.}
\vskip -0.2cm
\end{figure}

Band 1 is built on the $11/2^-$ state, suggesting that it has one proton in the lowest $h_{11/2}$ orbital with negative signature, which is largely favored relative to the positive-signature orbital, see Fig. \ref{fig1s}(a). In CNS notation the configuration is $[21_-;84]$ at low spin, and goes through a smooth alignment at $I$$\approx$18, which is most naturally understood as the $\nu (h_{11/2})^2$ alignment. The calculated band can be followed as a smooth terminating band to $I$=39.5, as seen in Fig. \ref{fig3}, where experiment and calculations are compared. This  $I$=39.5 state is the maximum spin of the $[21_-;84]$ configuration, where the observed band is seen up to two transitions short of termination.

Band 2 is observed for spin values $I$= 21.5$-$27.5 and is lower in energy than band 1. This might seem somewhat surprising, but gets a ready explanation from the CNS calculations as the $[21_-;A2]$ configuration, which has two less high-$j$ $h_{11/2}$ neutrons than $[21_-;84]$ and is calculated yrast above $I$$\approx$20, see Fig. \ref{fig1s}(b).
%Thus, the $N$=4 neutrons
%come closer to full $j$-shells, i.e. band 1 is built on the
%$\nu(gd)^{-6}(h_{11/2})^4$ configuration, while band 2 is built on the
%$\nu (gd)^{-4}(h_{11/2})^2$ configuration.
%It is thus somewhat similar to the terminating bands in the Er/Dy nuclei which start out with 6$-$8 $h_{11/2}$ protons, but terminates with a closed $Z$=64 core and only 2$-$4 $h_{11/2}$ protons.
%One would expect to see the neutron
%$h_{11/2}$ band crossing in $[21_-;A2]$ configuration assigned to band 2.
%However, because band 2 is only seen at high spin, it has already
%g%one through this band crossing when it is observed.
As noticed above, band 1 with four $h_{11/2}$ neutrons goes through a smooth crossing, in agreement with calculations, which only show a gradual variation of the slope in Fig. \ref{fig3}(b). In the $[21_-;A2]$ configuration assigned to band 2, aligned (or close-to-aligned) states are yrast at low energy for $I$=25.5, 27.5 and 29.5.  The exact spin values of the energetically favored configurations depend on the relative distance between the $g_{7/2}$ and $d_{5/2}$ shells, cf. Fig. \ref{fig5} below. With this in mind, it appears reasonable that band 2 is observed only to $I$=27.5, and not surprising that the relative energies of these states close to termination are not fully reproduced in the calculations.

%\begin{figure*}[!ht]
%\hskip -. cm
%\vskip -. cm
%\rotatebox{-0}{\scalebox{0.88}{\includegraphics[trim=10 0 0 0, width=\textwidth]{eimip-17.eps}}}
%\rotatebox{-0}{\scalebox{0.88}{\includegraphics[trim=10 0 0 0, width=\textwidth]{eimin-17.eps}}}
%\vskip -0.3 cm
%\caption {\label{fig6} Plots of $e_i$ versus $m_i$ for protons at $\varepsilon_2$=0.20, $\gamma$=60$^{\circ}$. For $N$=62, it is indicated how aligned states could be formed with 3 $h_{11/2}$ neutrons and five $(gd)$ neutron holes. Thus, a sloping Fermi surface is drawn through three $(gd)$ states. If two of these three states are empty, configurations with 5 $(gd)$ holes and three $h_{11/2}$ neurons are formed at $I_n$=19, 21 and 23.}
%\vskip -0.2cm
%\end{figure*}

\begin{figure}[!htb]
\hskip -. cm
\vskip -. cm
\rotatebox{-0}{\scalebox{0.4}{\includegraphics[trim=0 0 0 0, width=\textwidth]{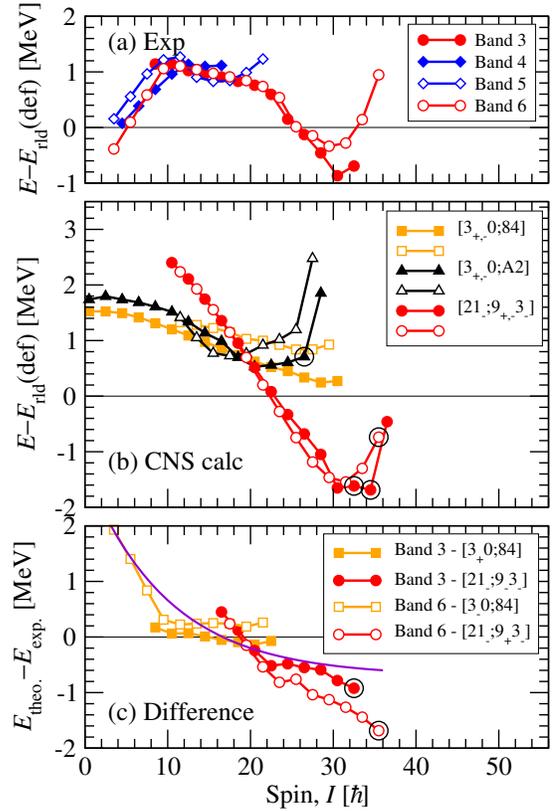}}}
\vskip -0.3 cm
\caption {\label{fig4} Same as Fig. \ref{fig3} but comparing of the experimental bands 3 and 6 with CNS configurations.}
\vskip -0.2cm
\end{figure}

%{\it Move further down: Fig. \ref{fig5} suggests that the configuration with maximum spin for the configurations involving $(gd)$ neutron holes are not very favored in energy, which means that most of the calculated bands become non-collective before they have reached their maximum spin values.}

\begin{figure}[!ht]
\hskip -. cm
\vskip -. cm
\rotatebox{-0}{\scalebox{0.48}{\includegraphics[trim=10 0 0 0, width=\textwidth]{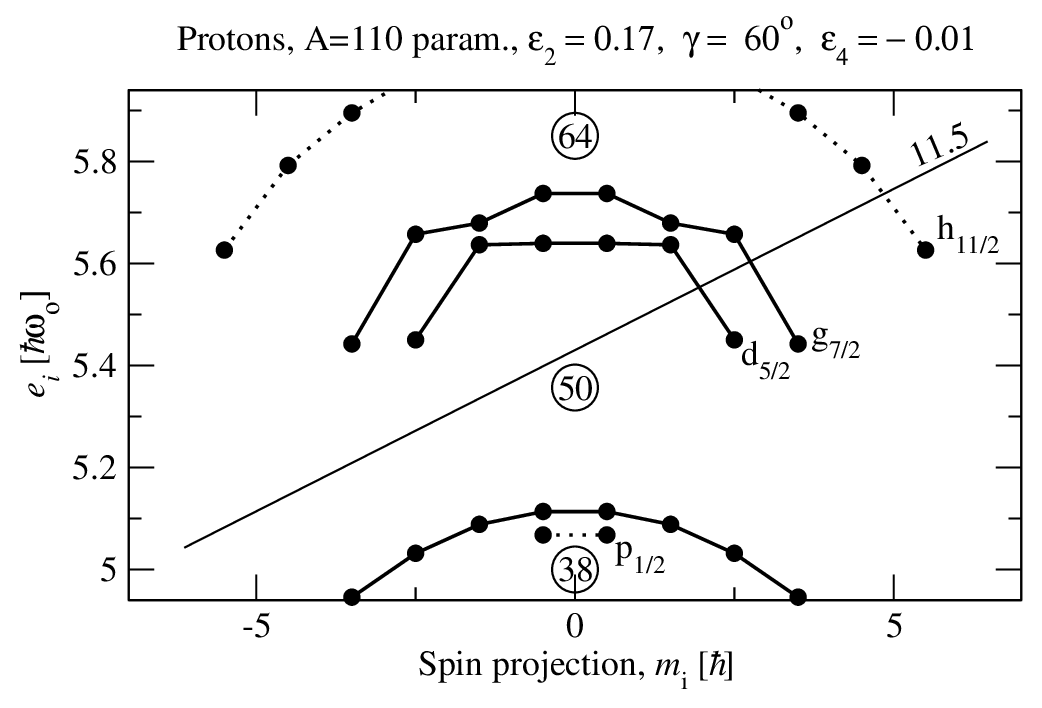}}}
\vskip 3.8 cm
\rotatebox{-0}{\scalebox{0.48}{\includegraphics[trim=10 0 0 0, width=\textwidth]{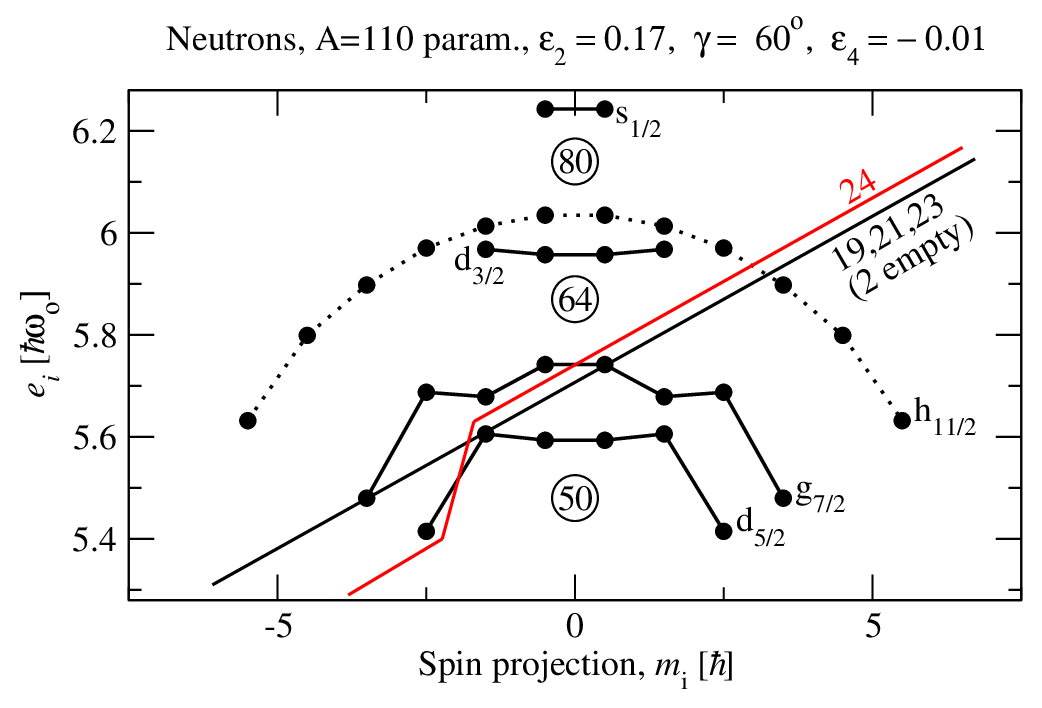}}}
\vskip -0.3 cm
\caption {\label{fig5} Plots of $e_i$ versus $m_i$ for protons and neutrons at $\varepsilon_2$=0.17, $\gamma$=60$^{\circ}$. For $Z=53$ it is shown how a favoured $I=11.5$ state
is formed with one $h_{11/2}$ and two $(dg)$ protons. For $N$=62, it is indicated how aligned states could be formed with 3 $h_{11/2}$ neutrons and five $(gd)$
neutron holes. Thus, a sloping Fermi surface is drawn through three $(gd)$ states. If two of these three states are empty, configurations with 5 $(gd)$ holes and three $h_{11/2}$ neurons are formed at $I_n$=19, 21 and 23. The red broken line shows how a maximum spin state for the same configuration but with signature $\alpha = -1/2$ is formed, i.e. for the terminating state of band 6.}
\vskip -0.2cm
\end{figure}

\begin{figure}[!htb]
\hskip -. cm
\vskip -. cm
\rotatebox{-0}{\scalebox{0.4}{\includegraphics[trim=0 0 0 0, width=\textwidth]{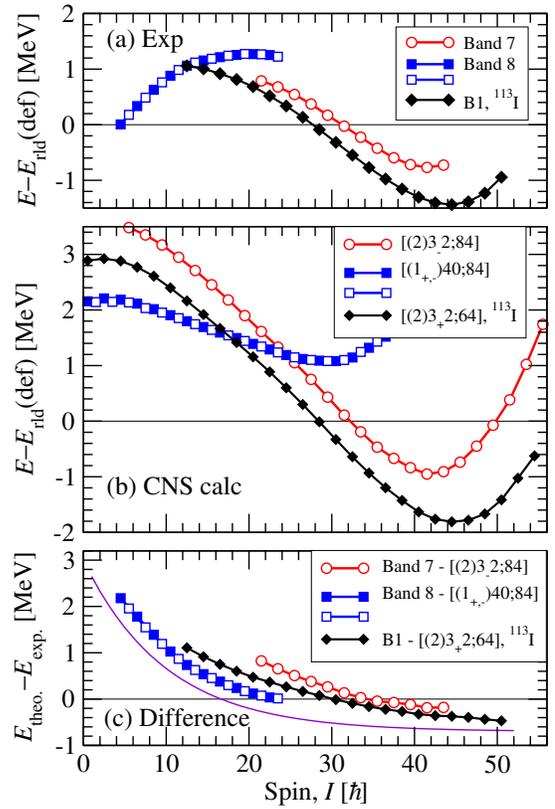}}}
\vskip -0.3 cm
\caption {\label{fig6} Same as Fig. \ref{fig3} but comparing of the experimental bands 7 and 8 with CNS configurations.}
\vskip -0.2cm
\end{figure}

\subsubsection{Bands 3$-$6}

Band 6 starts at $I$=7/2$^+$ at an excitation energy of 56 keV above the $5/2^+$ ground state.  As we have seen in Fig. \ref{fig1}, sharp crossings occur at $I$$\approx$10 in bands 4, 5 and 6, while band 3 is not really observed below the crossing which might also be sharp. This is different from the smooth $\nu (h_{11/2})^2$ alignments observed in band 1 at $I$$\approx$18 and in band 8 at $I$$\approx$10.

Here we will concentrate on bands 3 and 6 which are observed to highest spin values. These bands are assigned to $(dg)^3$ proton configurations combined with neutron configurations with an even number of neutrons in the positive- and negative-parity orbitals, i.e. $(dg)^8(h_{11/2})^4$ or $(dg)^{10}(h_{11/2})^2$ according to the discussion above. The energy curves of these configurations are drawn in the middle panel of Fig. \ref{fig4} where we note a crossing at $I$$\approx$15, where the $(h_{11/2})^4$ bands which are lower in energy at low spin, are crossed by the $(h_{11/2})^2$ bands which are lower at high spin. This observation suggests that the sharp crossings at $I$=10 are created by a change in configuration from a $(h_{11/2})^4$ configuration to a $(h_{11/2})^2$ configuration. 
%In general, with the $(h_{11/2})^4$ and $(h_{11/2})^2$ configurations close together, we should expect two pairs of bands based on the second lowest proton $g_{7/2}$ orbital 3/2$^+$[411]. This could explain the presence of the four bands 3-6 where 
Our interpretation is supported from the analysis of bands 1 and 2, showing that the relative energies of the $\nu(h_{11/2})^4$ and $\nu(h_{11/2})^2$ configurations are
more or less reproduced in our calculations. Furthermore, both in calculations, in general, and in experiment for bands 1 and 2, the $\nu(h_{11/2})^2$ configuration comes relatively lower in energy with increasing spin. Thus, the combination of these two neutron configurations with the two signatures of the proton in the $(dg)$ 3/2 [422] orbital will lead to two pairs of
bands which cross sharply somewhere around $I$=10$-$15. This is in general agreement with experiment, while a detailed understanding of their configurations could not  really be expected in the CNS formalism. 
% Alternatively, the four bands could be formed from the coupling of one neutron configuration to two different proton configurations with the odd proton in the second lowest and third lowest $h_{11/2}$ orbital.  However, the involvement of the $\nu(h_{11/2})^2$ configuration is supported by the fact that this configuration is assigned to band 2.
%Our conclusions
%are also in line with the assignments for bands 1,2 where the $(h_{11/2})^2$
%configuration comes below the $(h_{11/2})^4$ for $I \approx 20$ but the
%observed bands do never cross.
%with the present
%rather crude model, the agreement between experiment and calculations
%is clearly only qualitative but we cannot see any other plausible
%reason why the crossing in band 6 is so different from the crossing in
%band 1.
This interpretation is in line with general expectations that in high-$j$ shells, angular momentum is built more efficiently with fewer particles in the  shell. This is the case for the
proton $h_{11/2}$ shell in the $Z$=66$-$70 nuclei where with increasing spin, the number of $h_{11/2}$ protons for the yrast states decreases \cite{TB-IR-1985}, $(h_{11/2})^8 \rightarrow (h_{11/2})^6 \rightarrow (h_{11/2})^4$ or $(h_{11/2})^7 \rightarrow (h_{11/2})^5$, see also Fig. 12.11 in Ref. \cite{NR}.

%%%%%%%%%
%However, most often, the first crossing occurring in bands built on one-particle configurations, is interpreted as due to the alignment of a pair of high-$j$ particles, in our case of $h_{11/2}$ neutrons, see e. g. \cite{117I-warring,117I-paul} because the proton alignment is expected to occur at higher rotational frequency and lead to higher excited bands.  The present proposed interpretation challenges the more natural interpretation of the crossings at $I$=10 in bands 3$-$6 as induced by the alignment of two $h_{11/2}$ neutrons.  
%%%%%%%%%%

Crossings in high-$j$ shells are generally understood as so called 'paired crossings', i.e. a re-coupling of the spin vectors within the shell, see e. g. \cite{Steph73,NR,frau-1981}. This is well understood in the bottom of the shell. However, as shown in Ref. \cite{ing-phys-rep}, higher up in the shell when the pairing is not too strong, such crossings are naturally
understood as a decrease of the number of particles in the high-$j$ shell. This interpretation challenges the more standard interpretation of crossings like that in bands 3-6 in $^{115}$I. However, it offers an explanation to the fact that only a very smooth crossing is observed in band 1, while a much sharper crossing is seen in bands 3-6.

%The crossings at $I$$\approx$10 are naturally induced by the $\nu
%(h_{11/2})^2$ alignment. Then, above $I$$\approx$10, bands 3 and 6
%look like signature partners up to their highest spin values above
%¤$I$=30.

Starting from $I \approx 10$, bands 3 and 6 look like signature partners where they are assigned to a $(dg)^3$ proton configuration. They go through a second crossing at $I$$\approx$20.
%According to our interpretation,
%they are assigned
%to the $[3_{\pm}0;A2]$ configurations.
One would then expect that this crossing is caused by a pair transfer, but there is no such calculated band at low energy for the highest spin values, see Fig. \ref{fig1s}(a). Instead, all bands which are low in energy for $I$=20$-$40 have at least one proton in the $h_{11/2}$ shell. Indeed, the only reasonable interpretation of these bands which come down in energy around $I$=30 is the  $[21_-; 9_{\pm}3_-]$ configurations, which are clearly calculated as yrast in the spin range $I$=20$-$30, see Fig. \ref{fig1s}.

The experimental bands 3 and 6 are compared with calculations in Fig. \ref{fig4}. The difference curves show a decrease with spin, which is in line with general expectations for calculations without pairing, as indicated by the smooth line drawn in Fig. \ref{fig4}(c). The bands are observed up to spin values just above $I$=30, where the calculated $[21_{\pm};9_{\pm}3_{\pm}]$ bands tend to get non-collective with triaxiality sliding towards $\gamma$=60$^{\circ}$, with typical deformation $\varepsilon_2$=0.15$-$0.20. 

To get a better idea about how these non-collective states are built, one can make reference to the $e_i$ vs. $m_i$ diagrams in Fig. \ref{fig5}, where one can see that for $Z$=53, the aligned states which are clearly favored have $I_p$=11.5. For neutrons, these aligned states are best characterized by neutrons in the $h_{11/2}$ shell, but with neutron holes in the $(gd)$ shells which are more than half-full. As is illustrated in Fig. \ref{fig5}, relatively favored aligned states can be formed for $I_n$=19, 21, 23. Then, combining the proton and neutron spins results in favored states at $I$=30.5, 32.5 and 34.5, see Fig. \ref{fig5}. These calculated states are consistent with the highest spin states in band 3 with low-energy $I$=30.5 and 32.5 sates.

%These favored aligned states appear consistent with the total
%energy calculations, where the $I$=32.5 and 34.5 states are calculated
%low in energy and aligned, while the $I$=30.5 state is slightly
%collective with a shallow energy minimum at $\gamma$=52$^{\circ}$, see
%Fig. 2(a) in the supplementary material~\cite{supplemental}. 

Let us now consider the other signature for the [21;93] configuration corresponding to the observed band 6. It turns out that both in experiment and  calculations, the energy curve is smooth up to the highest spin in the configuration, $I=35.5$. The terminating state corresponds to
\[
\pi(dg)^2_6(h_{11/2})^1_{5.5} \nu(dg)^{-5}_{10.5}(h_{11/2})^3_{13.5}
\]
where the spin values in the different groups are specified in the subscripts. Note that this is the maximum spin of signature $\alpha$=$-1/2$ in the [21;93] configurations. To our knowledge, while many smooth terminating bands with holes in the proton core have been observed in this space \cite{IR-1995,ing-phys-rep}, this is the first time that such a band within the valence space has been observed up to its maximum spin value, $I_{max}$.

\subsubsection{Bands 7 and 8}

Band 7 starts at $I$=43/2$^+$ and is observed up to $I$=79/2$^+$. As the calculated close-to-yrast configurations without holes in the $g_{9/2}$ orbital have $E$$-$$E_{rld}$ minima at spins much lower than $I$=40, see Fig. \ref{fig1s}, the configuration of band 7 has to involve one or two holes in the $g_{9/2}$ orbital.  A configuration with one hole can be excluded because it would give rise to a strongly coupled band composed of dipole and cross-over quadrupole transitions, like in band 8. 

\begin{figure}[!ht]
\hskip -. cm
\vskip -. cm
\rotatebox{-0}{\scalebox{0.48}{\includegraphics[trim=10 0 0 0, width=\textwidth]{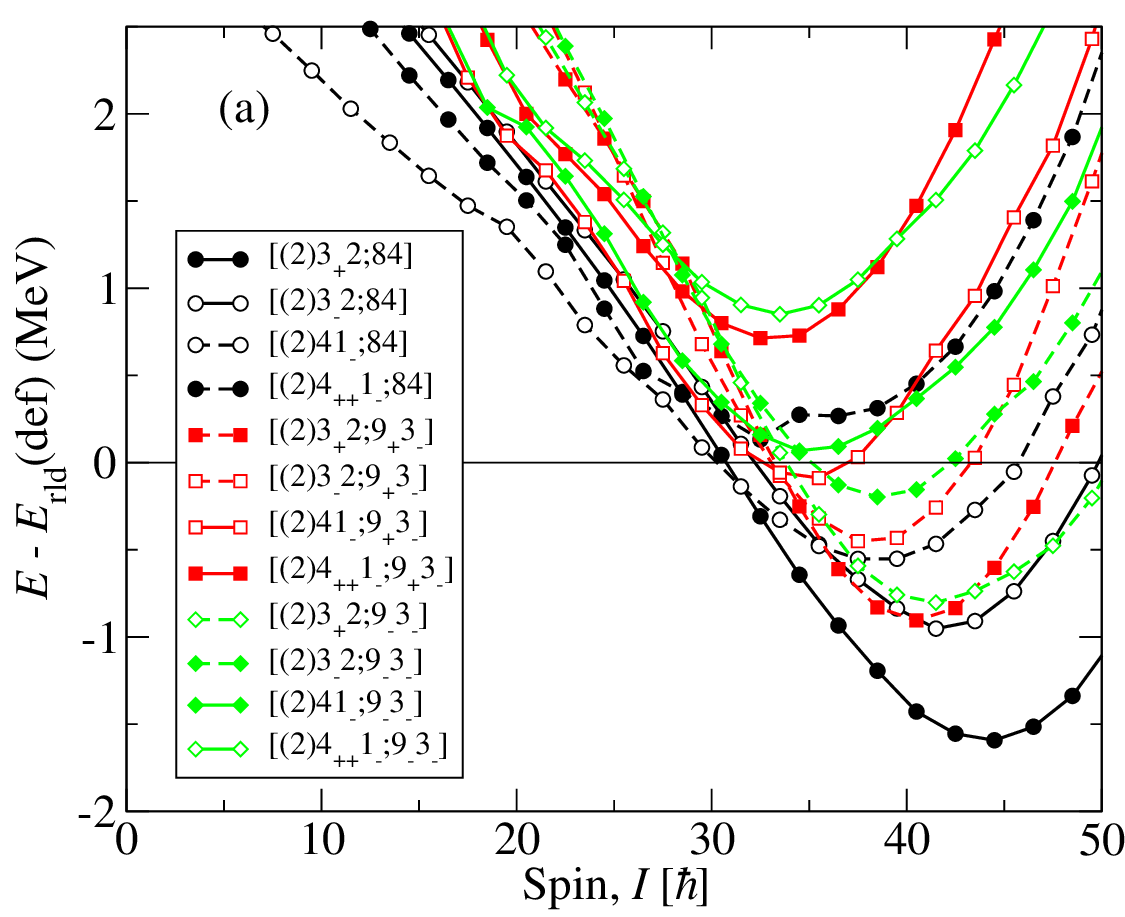}}}
\vskip 1.3 cm
\rotatebox{-0}{\scalebox{0.48}{\includegraphics[trim=10 0 0 0, width=\textwidth]{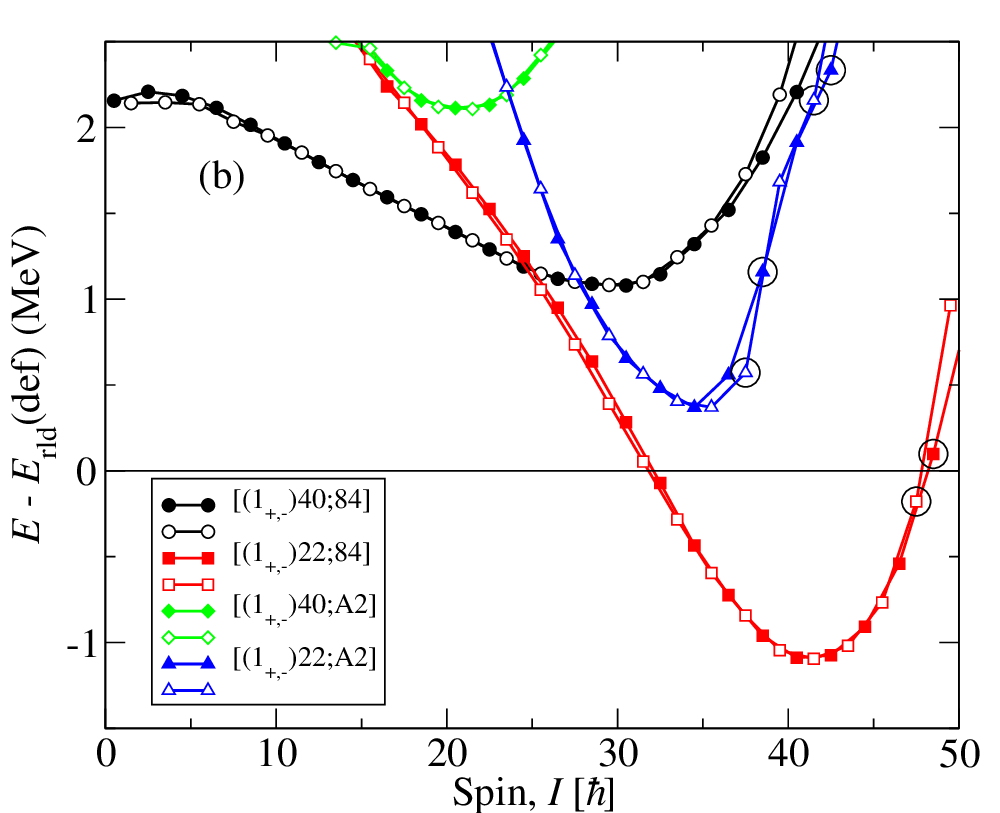}}}
\vskip -0.3 cm
\caption {\label{fig3s} (a) Low-energy configurations with two proton $g_{9/2}$ holes. (b) Low-energy configurations with one proton $g_{9/2}$ hole and an even number of particles in the other groups. }
\vskip -0.2cm
\end{figure}

We can thus conclude that band 7 has two holes in the $\pi g_{9/2}$ orbital.  The lowest energy configurations of this type are plotted in Fig. \ref{fig3s}(a). The most favored configuration for $I$$>$30 is $[(2)3_+2; 84]$ with signature $\alpha$=+1/2, which is different from the $\alpha$=$-1/2$ signature of band 7. However, the $[(2)3_-2; 84]$ signature partner is not much higher in energy for spin values up to $I$=40 and can be assigned to band 7.

Band 8 is strongly coupled and the two signatures remain degenerate up to the highest observed spin, $I$=23.5. The odd proton must be assigned to a high-$K$ orbital, and the only available such orbital is $g_{9/2}$, $9/2^+$[404]. This is consistent with the band head spin, $K$=9/2. With the odd proton in the $9/2^+$[404] orbital, an even number of protons and neutrons should be placed in the other groups. We can thus conclude that the additional protons must be distributed as either $(dg)^4$ or $(dg)^2(h_{11/2})^2$ while possible neutron configurations are $(dg)^8(h_{11/2})^4$ or $(dg)^{10}(h_{11/2})^2$. The possible combinations of these configurations are tested in Fig. \ref{fig3s}(b), from where one can conclude that for spin values up to $I$$\approx$25, [(1)40;84] bands are clearly lowest energy and should be assigned to band 8.

%$the positive. Thus, the additional protons
%must be and negative-parity
%orbitals, that is $[(1_{\pm})42; n_1n_2]$ configurations in CNS
%notation, as in many other nuclei of this mass region. Such
%configurations are plotted in Fig. 4(b) .

The bands 7 and 8 with two holes and one hole in the $g_{9/2}$ proton shell, respectively, are compared with their calculated counterpart in Fig. \ref{fig6}. The difference curves come out as expected with a smooth decrease with increasing spin. Indeed, if the $g_{9/2}$ orbital was lifted a few hundred keV, the difference curve would more or less coincide with the one which is used as an average for all bands of $^{115}$I, see Fig. \ref{fig6}(c). Note especially that while the $E$$-$$E_{rld}$ curves for band 8 in Fig. 6(a) and the configurations assigned to it in Fig. 6(b) do not show much resemblance, the difference curve in Fig. 6(c) comes out as expected considering that pairing is neglected in the calculations.
We have also included band 1 in $^{113}$I \cite{Sta01}. It is assigned to a configuration with two less $(dg)$ neutrons than the configuration of band 7 in $^{115}$I and leads to a very similar difference curve. This comparison illustrates the possibility to compare bands in different nuclei in the present CNS formalism \cite{Car06} and gives some extra confidence that our assignments for bands 7 and 8 are correct.

\section{Summary}

The level scheme of  $^{115}$I has been analyzed within the CNS formalism. With $Z$=53 and $N$=62, the low and intermediate spin properties are understood within the $(d_{5/2}g_{7/2})(h_{11/2})$ valence space outside the $Z$=$N$=50 core. The nucleus has enough particles above closed shells to form collective bands within the valence space, but not enough neutrons to render the excitations in $(d_{3/2}s_{1/2})$ orbitals competitive in energy. %Note also that in the CNS formalism, the $j$-shells are not the pure $j$-shells but those orbitals in the (deformation, rotation)-space which are dominated by these $j$-shells.
 The nucleus $^{115}$I exhibits a rich variety of configurations and coupling schemes when the proton particles outside the $Z=50$ are combined with neutron $h_{11/2}$ particles and $(dg)$ holes. %Note that the properties of the  $(dg)$ neutrons are best understood as holes because these $j$-shells are more than half-full.

%Within this valence space, configurations based on the
%$(dg)^2(h_{11/2})^1$ and the $(dg)^3$ configurations outside the $Z=50$ core. 
The bands based on the $\pi(h_{11/2})^1$ configuration are well understood when combined with neutron $(dg)^8(h_{11/2})^4$ and $(dg)^{10}(h_{11/2})^2$ configurations, leading to one smooth terminating band which is observed two transitions short of termination and another band which becomes non-collective before the
$I_{max}$-value is reached.

Also the positive-parity bands with all three protons in the $(dg)$ orbitals combine with the $(dg)^8(h_{11/2})^4$ and $(dg)^{10}(h_{11/2})^2$ neutron configurations. In this case, the details are more difficult to understand but it is suggested that the sharp crossings in the bands at $I$$\approx$$10$ can be understood as a crossing between $\nu(h_{11/2})^4$ and $\nu(h_{11/2})^2$ and configurations. At higher spin values above $I$$\approx$$20$ for these positive-parity bands, the  $(dg)^3$ proton configuration is not competitive, i.e. the $(dg)^2(h_{11/2})^1$ configuration takes over, combined with the $\nu(h_{11/2})^3$ configuration. In this case, the two signatures lead to one smooth terminating band and another band which becomes non-collective before the $I_{max}$-value is reached. This smooth terminating band is observed up to the maximum spin state in the configuration, i.e. to the $I_{max}$ state. While many such bands with holes in the $g_{9/2}$ orbitals have been observed for nuclei with mass numbers just below $^{115}$I \cite{IR-1995,ing-phys-rep}, this is to our knowledge the first time that such a smooth terminating band has been observed to termination within the present valence space. 

Going outside the valence space, the most competitive excitation is to lift protons from the $g_{9/2}$ shell of the $Z$=$50$ core. With one proton excited,  a strongly coupled band is formed in the $I$$\approx$5$-$25 spin range. With two protons excited, a smooth band with $E2$ transitions is observed for spin values $I$$\approx$20$-$45.  These bands are well understood in their full spin range within the CNS formalism, which suggests that they evolve towards termination pretty high above yrast.

%\section{Acknowledgments}

\bibliographystyle{apsrev4-1}

\bibliography{115I}

%merlin.mbs apsrev4-1.bst 2010-07-25 4.21a (PWD, AO, DPC) hacked
%Control: key (0)
%Control: author (72) initials jnrlst
%Control: editor formatted (1) identically to author
%Control: production of article title (-1) disabled
%Control: page (0) single
%Control: year (1) truncated
%Control: production of eprint (0) enabled
\begin{thebibliography}{22}%
\makeatletter
\providecommand \@ifxundefined [1]{%
 \@ifx{#1\undefined}
}%
\providecommand \@ifnum [1]{%
 \ifnum #1\expandafter \@firstoftwo
 \else \expandafter \@secondoftwo
 \fi
}%
\providecommand \@ifx [1]{%
 \ifx #1\expandafter \@firstoftwo
 \else \expandafter \@secondoftwo
 \fi
}%
\providecommand \natexlab [1]{#1}%
\providecommand \enquote  [1]{``#1''}%
\providecommand \bibnamefont  [1]{#1}%
\providecommand \bibfnamefont [1]{#1}%
\providecommand \citenamefont [1]{#1}%
\providecommand \href@noop [0]{\@secondoftwo}%
\providecommand \href [0]{\begingroup \@sanitize@url \@href}%
\providecommand \@href[1]{\@@startlink{#1}\@@href}%
\providecommand \@@href[1]{\endgroup#1\@@endlink}%
\providecommand \@sanitize@url [0]{\catcode `\\12\catcode `\$12\catcode
  `\&12\catcode `\#12\catcode `\^12\catcode `\_12\catcode `\%12\relax}%
\providecommand \@@startlink[1]{}%
\providecommand \@@endlink[0]{}%
\providecommand \url  [0]{\begingroup\@sanitize@url \@url }%
\providecommand \@url [1]{\endgroup\@href {#1}{\urlprefix }}%
\providecommand \urlprefix  [0]{URL }%
\providecommand \Eprint [0]{\href }%
\providecommand \doibase [0]{http://dx.doi.org/}%
\providecommand \selectlanguage [0]{\@gobble}%
\providecommand \bibinfo  [0]{\@secondoftwo}%
\providecommand \bibfield  [0]{\@secondoftwo}%
\providecommand \translation [1]{[#1]}%
\providecommand \BibitemOpen [0]{}%
\providecommand \bibitemStop [0]{}%
\providecommand \bibitemNoStop [0]{.\EOS\space}%
\providecommand \EOS [0]{\spacefactor3000\relax}%
\providecommand \BibitemShut  [1]{\csname bibitem#1\endcsname}%
\let\auto@bib@innerbib\@empty
%</preamble>
\bibitem [{\citenamefont {Bengtsson}\ and\ \citenamefont
  {Ragnarsson}(1985{\natexlab{a}})}]{Bengtsson1985}%
  \BibitemOpen
  \bibfield  {author} {\bibinfo {author} {\bibfnamefont {T.}~\bibnamefont
  {Bengtsson}}\ and\ \bibinfo {author} {\bibfnamefont {I.}~\bibnamefont
  {Ragnarsson}},\ }\href {\doibase 10.1016/0375-9474(85)90541-X} {\bibfield
  {journal} {\bibinfo  {journal} {Nucl. Phys. A}\ }\textbf {\bibinfo {volume}
  {436}},\ \bibinfo {pages} {14} (\bibinfo {year}
  {1985}{\natexlab{a}})}\BibitemShut {NoStop}%
\bibitem [{\citenamefont {Ragnarsson}\ \emph {et~al.}(2024)\citenamefont
  {Ragnarsson}, \citenamefont {Kardan}, \citenamefont {Carlsson}, \citenamefont
  {Paul}, \citenamefont {Petrache}, \citenamefont {Riley},\ and\ \citenamefont
  {Sharpey-Schafer}}]{Rag24}%
  \BibitemOpen
  \bibfield  {author} {\bibinfo {author} {\bibfnamefont {I.}~\bibnamefont
  {Ragnarsson}}, \bibinfo {author} {\bibfnamefont {A.}~\bibnamefont {Kardan}},
  \bibinfo {author} {\bibfnamefont {B.~G.}\ \bibnamefont {Carlsson}}, \bibinfo
  {author} {\bibfnamefont {E.~S.}\ \bibnamefont {Paul}}, \bibinfo {author}
  {\bibfnamefont {C.~M.}\ \bibnamefont {Petrache}}, \bibinfo {author}
  {\bibfnamefont {M.~A.}\ \bibnamefont {Riley}}, \ and\ \bibinfo {author}
  {\bibfnamefont {J.}~\bibnamefont {Sharpey-Schafer}, \bibfnamefont
  {J.~F.~Simpson}},\ }\href {\doibase 10.1103/PhysRevC.110.034313} {\bibfield
  {journal} {\bibinfo  {journal} {Phys. Rev. C}\ }\textbf {\bibinfo {volume}
  {110}},\ \bibinfo {pages} {034313} (\bibinfo {year} {2024})}\BibitemShut
  {NoStop}%
\bibitem [{\citenamefont {Stephens}\ \emph {et~al.}(1973)\citenamefont
  {Stephens}, \citenamefont {Diamond},\ and\ \citenamefont
  {Nilsson}}]{Steph73}%
  \BibitemOpen
  \bibfield  {author} {\bibinfo {author} {\bibfnamefont {F.~S.}\ \bibnamefont
  {Stephens}}, \bibinfo {author} {\bibfnamefont {R.~M.}\ \bibnamefont
  {Diamond}}, \ and\ \bibinfo {author} {\bibfnamefont {S.~G.}\ \bibnamefont
  {Nilsson}},\ }\href {\doibase 10.1016/0370-2693(73)90325-0} {\bibfield
  {journal} {\bibinfo  {journal} {Phys. Lett. B}\ }\textbf {\bibinfo {volume}
  {48}},\ \bibinfo {pages} {429} (\bibinfo {year} {1973})}\BibitemShut
  {NoStop}%
\bibitem [{\citenamefont {Nilsson}\ and\ \citenamefont {Ragnarsson}()}]{NR}%
  \BibitemOpen
  \bibfield  {author} {\bibinfo {author} {\bibfnamefont {S.~G.}\ \bibnamefont
  {Nilsson}}\ and\ \bibinfo {author} {\bibfnamefont {I.}~\bibnamefont
  {Ragnarsson}},\ }\href@noop {} {\bibinfo  {journal} {{\it Shapes and Shells
  in Nuclear Structure}, (Cambridge University), (1995)}\ }\BibitemShut
  {NoStop}%
\bibitem [{\citenamefont {Piel}\ \emph {et~al.}(1985)\citenamefont {Piel},
  \citenamefont {Chowdhury}, \citenamefont {Garg}, \citenamefont {Quader},
  \citenamefont {Stwertka}, \citenamefont {Vajda},\ and\ \citenamefont
  {Fossan}}]{115i-Piel}%
  \BibitemOpen
\bibfield  {journal} {  }\bibfield  {author} {\bibinfo {author} {\bibfnamefont
  {W.~F.~J.}\ \bibnamefont {Piel}}, \bibinfo {author} {\bibfnamefont
  {P.}~\bibnamefont {Chowdhury}}, \bibinfo {author} {\bibfnamefont
  {U.}~\bibnamefont {Garg}}, \bibinfo {author} {\bibfnamefont {M.~A.}\
  \bibnamefont {Quader}}, \bibinfo {author} {\bibfnamefont {P.~M.}\
  \bibnamefont {Stwertka}}, \bibinfo {author} {\bibfnamefont {S.}~\bibnamefont
  {Vajda}}, \ and\ \bibinfo {author} {\bibfnamefont {D.~B.}\ \bibnamefont
  {Fossan}},\ }\href {\doibase 10.1103/PhysRevC.31.456} {\bibfield  {journal}
  {\bibinfo  {journal} {Phys. Rev. C}\ }\textbf {\bibinfo {volume} {31}},\
  \bibinfo {pages} {456} (\bibinfo {year} {1985})}\BibitemShut {NoStop}%
\bibitem [{\citenamefont {Paul~{\it et al.}}(1992)}]{115i-Paul-Stony-Brook}%
  \BibitemOpen
  \bibfield  {author} {\bibinfo {author} {\bibfnamefont {E.~S.}\ \bibnamefont
  {Paul~{\it et al.}}},\ }\href {\doibase 10.1088/0954-3899/18/5/012}
  {\bibfield  {journal} {\bibinfo  {journal} {J. Phys. G: Nucl. Part. Phys.}\
  }\textbf {\bibinfo {volume} {18}},\ \bibinfo {pages} {837} (\bibinfo {year}
  {1992})}\BibitemShut {NoStop}%
\bibitem [{\citenamefont {Paul}\ \emph {et~al.}(1994)\citenamefont {Paul},
  \citenamefont {Andrews}, \citenamefont {Janzen}, \citenamefont {Radford},
  \citenamefont {Ward}, \citenamefont {Drake}, \citenamefont {DeGraaf},
  \citenamefont {Pilotte},\ and\ \citenamefont
  {Ragnarsson}}]{115i-Paul-Chalk-River}%
  \BibitemOpen
  \bibfield  {author} {\bibinfo {author} {\bibfnamefont {E.~S.}\ \bibnamefont
  {Paul}}, \bibinfo {author} {\bibfnamefont {H.~R.}\ \bibnamefont {Andrews}},
  \bibinfo {author} {\bibfnamefont {V.~P.}\ \bibnamefont {Janzen}}, \bibinfo
  {author} {\bibfnamefont {D.~C.}\ \bibnamefont {Radford}}, \bibinfo {author}
  {\bibfnamefont {D.}~\bibnamefont {Ward}}, \bibinfo {author} {\bibfnamefont
  {T.~E.}\ \bibnamefont {Drake}}, \bibinfo {author} {\bibfnamefont
  {J.}~\bibnamefont {DeGraaf}}, \bibinfo {author} {\bibfnamefont
  {S.}~\bibnamefont {Pilotte}}, \ and\ \bibinfo {author} {\bibfnamefont
  {I.}~\bibnamefont {Ragnarsson}},\ }\href {\doibase 10.1103/PhysRevC.150.741}
  {\bibfield  {journal} {\bibinfo  {journal} {Phys. Rev. C}\ }\textbf {\bibinfo
  {volume} {50}},\ \bibinfo {pages} {741} (\bibinfo {year} {1994})}\BibitemShut
  {NoStop}%
\bibitem [{\citenamefont {Moon}\ \emph {et~al.}(2003)\citenamefont {Moon},
  \citenamefont {Komatsubara},\ and\ \citenamefont {Furuno}}]{115i-Moon}%
  \BibitemOpen
  \bibfield  {author} {\bibinfo {author} {\bibfnamefont {C.-B.}\ \bibnamefont
  {Moon}}, \bibinfo {author} {\bibfnamefont {T.}~\bibnamefont {Komatsubara}}, \
  and\ \bibinfo {author} {\bibfnamefont {K.}~\bibnamefont {Furuno}},\
  }\href@noop {} {\bibfield  {journal} {\bibinfo  {journal} {Journal of Korean
  Physical Society}\ }\textbf {\bibinfo {volume} {43}},\ \bibinfo {pages} {319}
  (\bibinfo {year} {2003})}\BibitemShut {NoStop}%
\bibitem [{\citenamefont {Pakarinen~{\it et al.}}(2020)}]{jurogam3}%
  \BibitemOpen
  \bibfield  {author} {\bibinfo {author} {\bibfnamefont {J.}~\bibnamefont
  {Pakarinen~{\it et al.}}},\ }\href {\doibase 10.1140/epja/s10050-020-00144-6}
  {\bibfield  {journal} {\bibinfo  {journal} {Eur. Phys. J. A}\ }\textbf
  {\bibinfo {volume} {56}},\ \bibinfo {pages} {149} (\bibinfo {year}
  {2020})}\BibitemShut {NoStop}%
\bibitem [{\citenamefont {Jodidar}\ and\ \citenamefont
  {Petrache}()}]{115I-exp}%
  \BibitemOpen
  \bibfield  {author} {\bibinfo {author} {\bibfnamefont {P.~M.}\ \bibnamefont
  {Jodidar}}\ and\ \bibinfo {author} {\bibfnamefont {C.~M.}\ \bibnamefont
  {Petrache}},\ }\href@noop {} {\bibinfo  {journal} {{\it et al.}, to be
  published}\ }\BibitemShut {NoStop}%
\bibitem [{\citenamefont {Jodidar~{\it et al.}}(2024)}]{117cs-jodidar}%
  \BibitemOpen
\bibfield  {journal} {  }\bibfield  {author} {\bibinfo {author} {\bibfnamefont
  {P.~M.}\ \bibnamefont {Jodidar~{\it et al.}}},\ }\href {\doibase
  10.1103/PhysRevC.110.044304} {\bibfield  {journal} {\bibinfo  {journal}
  {Phys. Rev. C}\ }\textbf {\bibinfo {volume} {110}},\ \bibinfo {pages}
  {044304} (\bibinfo {year} {2024})}\BibitemShut {NoStop}%
\bibitem [{\citenamefont {Evans~{\it et al.}}(2006)}]{evans}%
  \BibitemOpen
  \bibfield  {author} {\bibinfo {author} {\bibfnamefont {A.~O.}\ \bibnamefont
  {Evans~{\it et al.}}},\ }\href {\doibase 10.1016/j.physletb.2006.03.020}
  {\bibfield  {journal} {\bibinfo  {journal} {Phys. Lett. B}\ }\textbf
  {\bibinfo {volume} {636}},\ \bibinfo {pages} {25} (\bibinfo {year}
  {2006})}\BibitemShut {NoStop}%
\bibitem [{\citenamefont {Warring~{\it et al.}}(1993)}]{117I-warring}%
  \BibitemOpen
  \bibfield  {author} {\bibinfo {author} {\bibfnamefont {M.~P.}\ \bibnamefont
  {Warring~{\it et al.}}},\ }\href {\doibase 10.1103/PhysRevC.48.2629}
  {\bibfield  {journal} {\bibinfo  {journal} {Phys. Rev. C}\ }\textbf {\bibinfo
  {volume} {48}},\ \bibinfo {pages} {2629} (\bibinfo {year}
  {1993})}\BibitemShut {NoStop}%
\bibitem [{\citenamefont {Starosta~{\it et al.}}(2001)}]{Sta01}%
  \BibitemOpen
  \bibfield  {author} {\bibinfo {author} {\bibfnamefont {K.}~\bibnamefont
  {Starosta~{\it et al.}}},\ }\href {\doibase 10.1103/PhysRevC.64.014304}
  {\bibfield  {journal} {\bibinfo  {journal} {Phys. Rev. C}\ }\textbf {\bibinfo
  {volume} {64}},\ \bibinfo {pages} {014304} (\bibinfo {year}
  {2001})}\BibitemShut {NoStop}%
\bibitem [{\citenamefont {Zheng~{\it et al.}}(2021)}]{119cs-zheng}%
  \BibitemOpen
  \bibfield  {author} {\bibinfo {author} {\bibfnamefont {K.~K.}\ \bibnamefont
  {Zheng~{\it et al.}}},\ }\href {\doibase 10.1103/PhysRevC.104.044305}
  {\bibfield  {journal} {\bibinfo  {journal} {Phys. Rev. C}\ }\textbf {\bibinfo
  {volume} {104}},\ \bibinfo {pages} {044305} (\bibinfo {year}
  {2021})}\BibitemShut {NoStop}%
\bibitem [{\citenamefont {Jodidar}\ and\ \citenamefont
  {Petrache}(2025)}]{114i-jodidar}%
  \BibitemOpen
  \bibfield  {author} {\bibinfo {author} {\bibfnamefont {P.~M.}\ \bibnamefont
  {Jodidar}}\ and\ \bibinfo {author} {\bibfnamefont {{\it et al.}.~C.~M.}\
  \bibnamefont {Petrache}},\ }\href {\doibase 10.1103/PhysRevC.111.014303}
  {\bibfield  {journal} {\bibinfo  {journal} {Phys. Rev. C}\ }\textbf {\bibinfo
  {volume} {111}},\ \bibinfo {pages} {104303} (\bibinfo {year}
  {2025})}\BibitemShut {NoStop}%
\bibitem [{\citenamefont {Ragnarsson}\ \emph {et~al.}(1995)\citenamefont
  {Ragnarsson}, \citenamefont {Janzen}, \citenamefont {Fossan}, \citenamefont
  {Schmeing},\ and\ \citenamefont {Wadsworth}}]{IR-1995}%
  \BibitemOpen
  \bibfield  {author} {\bibinfo {author} {\bibfnamefont {I.}~\bibnamefont
  {Ragnarsson}}, \bibinfo {author} {\bibfnamefont {V.~P.}\ \bibnamefont
  {Janzen}}, \bibinfo {author} {\bibfnamefont {D.~B.}\ \bibnamefont {Fossan}},
  \bibinfo {author} {\bibfnamefont {N.~C.}\ \bibnamefont {Schmeing}}, \ and\
  \bibinfo {author} {\bibfnamefont {R.}~\bibnamefont {Wadsworth}},\ }\href
  {\doibase 10.1103/PhysRevLett.74.3935} {\bibfield  {journal} {\bibinfo
  {journal} {Phys. Rev. Lett.}\ }\textbf {\bibinfo {volume} {74}},\ \bibinfo
  {pages} {3935} (\bibinfo {year} {1995})}\BibitemShut {NoStop}%
\bibitem [{\citenamefont {Afanasjev}\ \emph {et~al.}(1999)\citenamefont
  {Afanasjev}, \citenamefont {Fossan}, \citenamefont {Lane},\ and\
  \citenamefont {Ragnarsson}}]{ing-phys-rep}%
  \BibitemOpen
  \bibfield  {author} {\bibinfo {author} {\bibfnamefont {A.}~\bibnamefont
  {Afanasjev}}, \bibinfo {author} {\bibfnamefont {D.}~\bibnamefont {Fossan}},
  \bibinfo {author} {\bibfnamefont {G.}~\bibnamefont {Lane}}, \ and\ \bibinfo
  {author} {\bibfnamefont {I.}~\bibnamefont {Ragnarsson}},\ }\href {\doibase
  http://dx.doi.org/10.1016/S0370-1573(99)00035-6} {\bibfield  {journal}
  {\bibinfo  {journal} {Phys. Rep.}\ }\textbf {\bibinfo {volume} {322}},\
  \bibinfo {pages} {1 } (\bibinfo {year} {1999})}\BibitemShut {NoStop}%
\bibitem [{\citenamefont {Carlsson}\ and\ \citenamefont
  {Ragnarsson}(2006)}]{Car06}%
  \BibitemOpen
  \bibfield  {author} {\bibinfo {author} {\bibfnamefont {B.~G.}\ \bibnamefont
  {Carlsson}}\ and\ \bibinfo {author} {\bibfnamefont {I.}~\bibnamefont
  {Ragnarsson}},\ }\href {\doibase 10.1103/PhysRevC.74.011302} {\bibfield
  {journal} {\bibinfo  {journal} {Phys. Rev. C}\ }\textbf {\bibinfo {volume}
  {74}},\ \bibinfo {pages} {011302(R)} (\bibinfo {year} {2006})}\BibitemShut
  {NoStop}%
\bibitem [{\citenamefont {Zhang}\ \emph {et~al.}(1989)\citenamefont {Zhang},
  \citenamefont {Xu}, \citenamefont {Fossan}, \citenamefont {Liang},
  \citenamefont {Ma},\ and\ \citenamefont {Paul}}]{A=120-140}%
  \BibitemOpen
  \bibfield  {author} {\bibinfo {author} {\bibfnamefont {J.-y.}\ \bibnamefont
  {Zhang}}, \bibinfo {author} {\bibfnamefont {N.}~\bibnamefont {Xu}}, \bibinfo
  {author} {\bibfnamefont {D.~B.}\ \bibnamefont {Fossan}}, \bibinfo {author}
  {\bibfnamefont {Y.}~\bibnamefont {Liang}}, \bibinfo {author} {\bibfnamefont
  {R.}~\bibnamefont {Ma}}, \ and\ \bibinfo {author} {\bibfnamefont {E.~S.}\
  \bibnamefont {Paul}},\ }\href {\doibase 10.1103/PhysRevC.39.714} {\bibfield
  {journal} {\bibinfo  {journal} {Phys. Rev. C}\ }\textbf {\bibinfo {volume}
  {39}},\ \bibinfo {pages} {714} (\bibinfo {year} {1989})}\BibitemShut
  {NoStop}%
\bibitem [{\citenamefont {Bengtsson}\ and\ \citenamefont
  {Ragnarsson}(1985{\natexlab{b}})}]{TB-IR-1985}%
  \BibitemOpen
  \bibfield  {author} {\bibinfo {author} {\bibfnamefont {T.}~\bibnamefont
  {Bengtsson}}\ and\ \bibinfo {author} {\bibfnamefont {I.}~\bibnamefont
  {Ragnarsson}},\ }\href {\doibase 10.1016/0370-2693(85)90186-8} {\bibfield
  {journal} {\bibinfo  {journal} {Phys. Lett. B}\ }\textbf {\bibinfo {volume}
  {163}},\ \bibinfo {pages} {31} (\bibinfo {year}
  {1985}{\natexlab{b}})}\BibitemShut {NoStop}%
\bibitem [{\citenamefont {Frauendorf}(1981)}]{frau-1981}%
  \BibitemOpen
  \bibfield  {author} {\bibinfo {author} {\bibfnamefont {S.}~\bibnamefont
  {Frauendorf}},\ }\href {\doibase 10.1088/0031-8949/24/1B/034} {\bibfield
  {journal} {\bibinfo  {journal} {Phys. Scr.}\ }\textbf {\bibinfo {volume}
  {24}},\ \bibinfo {pages} {349} (\bibinfo {year} {1981})}\BibitemShut
  {NoStop}%
\end{thebibliography}%
%\clearpage

\end{document}